\documentclass[10pt]{article}
\usepackage{amssymb}
\usepackage{amsmath}
\usepackage{amsthm}
\usepackage{color}
\usepackage{comment}
\usepackage{geometry}		
\usepackage{graphicx}
\usepackage{siunitx}
\usepackage{url}

\let\originalleft\left
\let\originalright\right
\renewcommand{\left}{\mathopen{}\mathclose\bgroup\originalleft}
\renewcommand{\right}{\aftergroup\egroup\originalright}

\usepackage{tabularx,booktabs}
\usepackage{caption}
\usepackage{subcaption}

\usepackage{hyperref}
\hypersetup{
    colorlinks=true,
    linkcolor=blue,
    urlcolor=blue,
    citecolor=blue
}

\usepackage{tikz}
\usepackage{array}
\usepackage{booktabs}
\usepackage{epstopdf}

\usepackage{fancyhdr}

\usepackage{mathtools}
\usepackage{tensor}

\theoremstyle{definition}

\theoremstyle{remark}

\usepackage{authblk} 
\usepackage{orcidlink} 
\usepackage{lineno}

\usetikzlibrary{arrows.meta, positioning}

\title{Short-term associations between wild-bird and mammalian high pathogenic avian influenza detection in the United States}

\author[1]{H.O.~Fatoyinbo\,\orcidlink{0000-0002-6036-2957}\thanks{Corresponding author: \href{mailto:hammed.fatoyinbo@aut.ac.nz}{hammed.fatoyinbo@aut.ac.nz}}}

\affil[1]{Department of Mathematical Sciences, Auckland University of Technology, Auckland 1010, New Zealand}

\date{\today}

\begin{document}

\maketitle

\begin{abstract}
    High pathogenic avian influenza (HPAI) detections in mammals have raised concern about repeated wildlife
spillover and the targeting of mammalian surveillance. We evaluated whether wild-bird detection records
contained short-term explanatory and predictive information about mammalian detections in the United
States. Public surveillance records from April 2022 to July 2026 comprised 18,523 wild-bird and 815 mammalian
detections. Monthly state-level mammalian counts were analysed using negative-binomial mixed
models adjusted for seasonality, preceding mammalian detections and state-level heterogeneity. The selected
model contained contemporaneous and one-month-lagged same-state wild-bird counts. A doubling
of one plus the contemporaneous wild-bird count was associated with a 57.9\% increase in the expected
mammalian count (95\% CI 43.5–73.8\%), and a doubling of the preceding-month count with a 15.9\% increase
(5.2–27.8\%). Bordering-state exposure did not improve fit. In a six-month temporal holdout, adding
same-state wild-bird exposure reduced mean absolute error by 23.8\% and root mean squared error by 27.9\%,
although total detections remained overpredicted. After exact-row deduplication, the corresponding effects
were 59.7\% and 13.3\%. Wild-bird surveillance records contained short-term, geographically local information
about mammalian HPAI detection activity, but the associations do not establish direct transmission and
require further calibration before operational forecasting.
\end{abstract}

\newpage

\section{Introduction}
The continuing panzootic of high pathogenic avian influenza (HPAI) A(H5N1), particularly clade 2.3.4.4b viruses, has altered disease ecology at the wildlife--domestic animal interface. Since 2020, sustained circulation in wild birds has been accompanied by intercontinental spread, reassortment and an expanding range of infections in terrestrial and marine mammals \cite{elsmo2023,plaza2024,koopmans2024,peacock2025,puryear2024wildlife,damodaran2026}. The contemporary panzootic differs from earlier H5N1 waves in its geographical reach, impact on wild birds and diversity of affected mammalian hosts \cite{plaza2024,puryear2024wildlife,peacock2025,enetwild2024mammals}.

Mammalian detections indicate repeated opportunities for cross-species transmission, may involve severe neurological or respiratory disease, and create settings in which viral adaptation or onward transmission could occur \cite{elsmo2023,puryear2024wildlife,barbachano2023spillover,peacock2025}. Avian influenza viruses may become better adapted to mammalian hosts through reassortment or through mutation and selection following direct spillover and subsequent transmission \cite{barbachano2023spillover}. Transmission within United States dairy cattle further demonstrated that an avian-origin virus can enter and circulate within a managed mammalian population \cite{koopmans2024,peacock2025,usda2026}. Integrated surveillance across wildlife, livestock and environmental sectors is therefore central to contemporary One Health preparedness \cite{koopmans2024,woah2026}.

Wild birds are central to the maintenance and geographical dissemination of contemporary H5 viruses, but pathways to mammals are heterogeneous. Exposure may occur through predation, scavenging, consumption of infected birds, contact with contaminated water or substrates, or use of shared habitats \cite{wood2010,root2020synanthropic,elsmo2023,enetwild2024mammals}. A systematic review found that reported wild-mammal infections occurred most often in carnivores and were associated principally with predation, scavenging or contact with infected birds, while evidence for sustained transmission within free-ranging mammalian populations remained limited \cite{enetwild2024mammals}. Mammal-to-mammal transmission may nevertheless contribute in particular settings, including large outbreaks in pinnipeds \cite{uhart2024elephantseals,peacock2025}. Temporal alignment between surveillance streams can therefore reflect avian exposure, continuing transmission in mammals, shared environmental drivers, intensified surveillance, or several processes together. Such alignment is not proof of direct bird-to-mammal transmission.

Wild-bird surveillance may nevertheless provide an accessible indicator for prioritising mammalian sampling. Rapid hunter-harvest surveillance has detected H5N1 reproducibly across locations and has been proposed as an early-warning contribution \cite{shemmingspayne2024}. Long-term sampling also shows that avian influenza occurrence varies by host, place and year \cite{stanislawek2024}. Surveillance-based early-warning models have detected seasonal shifts in HPAI transmission, and recent forecasting studies have shown that integrating multiple information streams can improve predictive performance \cite{hill2023,movahedinia2026}. Recent United States work also demonstrates strong species-specific and regional clustering at the wild-waterfowl--poultry interface \cite{varga2026}.

However, raw co-occurrence is affected by seasonality, persistent local events, geographical heterogeneity, reporting delay and sampling effort. National correlations cannot establish whether associations remain after adjustment, whether the informative signal is local or spatially diffuse, or whether explanatory associations improve out-of-sample prediction. Cross-species HPAI modelling remains comparatively limited \cite{wang2026}. Recent studies have begun to combine ecological risk factors, surveillance records and mechanistic transmission models across wildlife, poultry and dairy cattle, but cross-host models with prospective validation remain uncommon \cite{wang2026,canada2026spatiotemporal,tiwari2025stochastic}. We therefore characterised both surveillance streams, selected a same-state exposure window, tested bordering-state exposure, evaluated temporal holdout performance and assessed robustness to exact-row deduplication. We hypothesised that elevated same-state bird activity would be associated with higher mammalian counts over a short window and would improve prediction. Analyses estimate record-level associations, not incidence, individual spillover probability or causal transmission.

\section{Materials and methods}
\subsection{Data sources and preparation}
We analysed publicly available United States Department of Agriculture Animal and Plant Health Inspection Service wild-bird and mammalian detection records \cite{usda2026}. The primary period comprised complete months from 1 April 2022 to 31 July 2026. Collection date was used as the event date. Records were standardised and aggregated to a balanced state-month panel. Missing state-month combinations were assigned zero detections; these zeros indicated no recorded positive detection, not confirmed absence of infection or surveillance. Exact repeated rows were retained in the primary analysis because unique specimen identifiers were unavailable and were removed in a sensitivity analysis.

\subsection{Statistical analysis}
Let $B_{i,t}$ and $M_{i,t}$ denote the numbers of wild-bird and mammalian detection records in state $i$ during month $t$. Mammalian counts were analysed using generalized linear mixed models with a log link and state-specific random intercept, an approach suited to non-Gaussian repeated count data with between-state heterogeneity \cite{bolker2009glmm}. Poisson, negative-binomial type 1 (NB1), negative-binomial type 2 (NB2), and zero-inflated NB2 distributions were compared on common observations using information criteria, convergence checks, simulated residual diagnostics and zero-count reproduction \cite{burnham2002model,brooks2017,warton2005zeros,hartig2026dharma}. A high proportion of zero counts was not, by itself, treated as evidence that a structural zero-inflation process was required \cite{warton2005zeros}.

The baseline linear predictor was
\begin{equation}
\log(\mu_{i,t})=\alpha+u_i+\gamma\log(1+M_{i,t-1})+a_1\sin\!\left(\frac{2\pi t}{12}\right)+a_2\cos\!\left(\frac{2\pi t}{12}\right),
\label{eq:baseline}
\end{equation}
where $\mu_{i,t}=E(M_{i,t}\mid\mathbf{x}_{i,t},u_i)$, $\alpha$ is the intercept, $u_i\sim N(0,\sigma_u^2)$ is the state random intercept, $\gamma$ represents short-term persistence in mammalian detections, and the sine and cosine terms represent annual seasonality.

Candidate same-state models represented the exposure--lag relationship by including log-transformed wild-bird counts from the current and preceding months. This follows the general principle of distributed-lag modelling, in which an exposure may have associations extending across several subsequent time intervals \cite{gasparrini2010distributed}. For a distributed-lag window ending at lag $L$, the model was
\begin{equation}
\log(\mu_{i,t})=\alpha+u_i+\gamma\log(1+M_{i,t-1})
+a_1\sin\!\left(\frac{2\pi t}{12}\right)+a_2\cos\!\left(\frac{2\pi t}{12}\right)
+\sum_{\ell=0}^{L}\beta_{\ell}\log(1+B_{i,t-\ell}).
\label{eq:lagmodel}
\end{equation}
We compared a no-bird model, individual lags 0--3, and distributed windows containing lags 0--1, 0--2 and 0--3. Model comparison used AIC, Akaike weights, likelihood-ratio tests, coefficient stability and simulated residual behaviour; AIC differences and weights were interpreted within an information-theoretic framework \cite{burnham2002model}. The percentage change associated with doubling one plus a count predictor was
\begin{equation}
\Delta_{\ell}(\%)=100\left(2^{\beta_{\ell}}-1\right).
\label{eq:doubling}
\end{equation}
For a joint change in several lagged exposures, the corresponding contrast is $100\{2^{\sum_{\ell\in\mathcal L}\beta_\ell}-1\}$.

For the spatial extension, neighbouring-state exposure was constructed from a binary contiguity matrix with row-standardised weights, so that exposure represented the mean wild-bird count among states sharing a boundary \cite{anselin1988spatial}:
\begin{equation}
N_{i,t}=\sum_{j\ne i}w_{ij}B_{j,t},\qquad
w_{ij}=\frac{A_{ij}}{\sum_{k\ne i}A_{ik}}.
\label{eq:neighbor}
\end{equation}
Here $A_{ij}=1$ when states $i$ and $j$ shared a boundary and $A_{ij}=0$ otherwise. The selected same-state model was compared with models adding $\log(1+N_{i,t})$, $\log(1+N_{i,t-1})$, or both. Alaska and Hawaii were excluded because they had no neighbours under the shared-border definition.

February--July 2026 was held out for prediction. A temporally ordered holdout was used rather than a random split to preserve the forecasting information structure and prevent future observations from entering model estimation \cite{tashman2000forecast}. The selected model was compared with the autoregressive seasonal baseline. Predictions used the observed preceding mammalian count and therefore represented conditional one-month-ahead predictions rather than a recursive six-month forecast. For test observations $k=1,\ldots,n$, with observed count $y_k$ and predictive mean $\widehat\mu_k$, point accuracy was summarised as
\begin{equation}
\operatorname{MAE}=\frac{1}{n}\sum_{k=1}^{n}|y_k-\widehat\mu_k|,
\label{eq:mae}
\end{equation}
\begin{equation}
\operatorname{RMSE}=\left\{\frac{1}{n}\sum_{k=1}^{n}(y_k-\widehat\mu_k)^2\right\}^{1/2}.
\label{eq:rmse}
\end{equation}
Probabilistic performance was assessed using mean negative log predictive density (NLPD),
\begin{equation}
\operatorname{NLPD}=-\frac{1}{n}\sum_{k=1}^{n}\log p\!\left(y_k\mid\widehat{\boldsymbol\theta}_k\right),
\label{eq:nlpd}
\end{equation}
where $p(y_k\mid\widehat{\boldsymbol\theta}_k)$ is the fitted NB2 predictive probability. Mean NLPD is a proper score for the full predictive distribution, while empirical prediction-interval coverage evaluates calibration of nominal intervals \cite{gneiting2007scoring,gneiting2007calibration}. Empirical coverage of nominal $100(1-\alpha)\%$ prediction intervals was
\begin{equation}
\widehat C_{1-\alpha}=\frac{1}{n}\sum_{k=1}^{n}I\!\left(L_{k,\alpha}\le y_k\le U_{k,\alpha}\right).
\label{eq:coverage}
\end{equation}
Analyses used R 4.5.2 with \texttt{glmmTMB} for mixed count models \cite{brooks2017}, \texttt{DHARMa} for simulation-based residual diagnostics \cite{hartig2026dharma}, and \texttt{sf} and \texttt{spData} for spatial data and contiguity structures.

\section{Results}
\subsection{Record volume, timing and geographical coverage}
The source files contained 19,978 wild-bird and 825 mammalian records. The primary period retained 18,523 wild-bird records from 248 species, 51 states or jurisdictions and 1,561 state--county combinations, and 815 mammalian records from 56 species, 42 states and 297 state--county combinations. Median non-negative collection-to-detection delay was 15 days in both streams. Exact repeated rows comprised 35.6\% of primary-period bird records and 25.3\% of mammalian records.

\begin{figure*}[htbp]
\centering
\includegraphics[width=0.6\columnwidth]{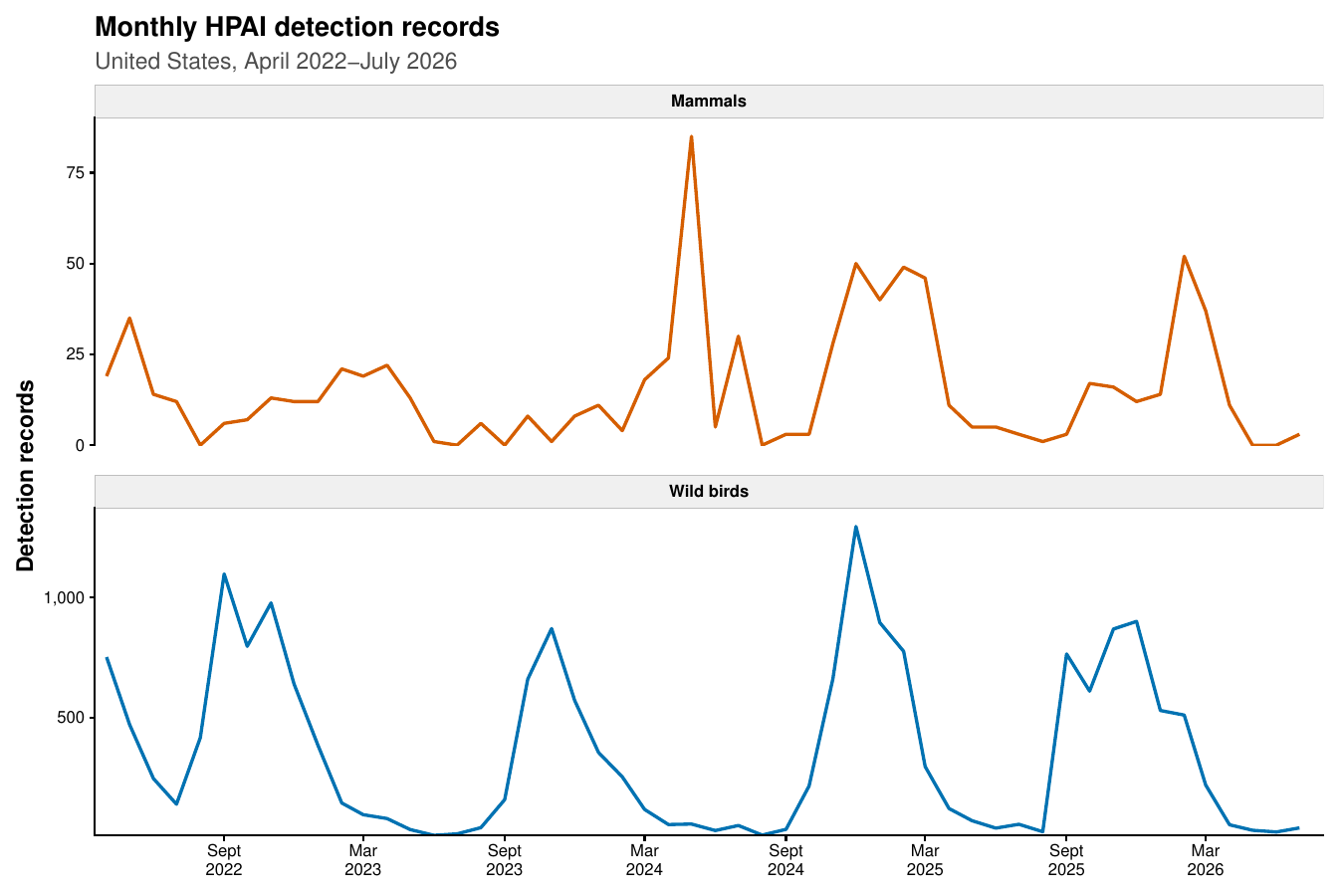}
\caption{Monthly HPAI detection records in wild birds and mammals, April 2022--July 2026. Counts are positive surveillance records, not incidence or prevalence.}
\label{fig:monthly}
\end{figure*}

The balanced panel contained 2,652 state-months. Mammalian counts had mean 0.307 and variance 4.515, a variance-to-mean ratio of 14.7. Zero records occurred in 89.9\% of state-months, while 268 state-months contained at least one mammalian detection. Of these positive state-months, 215 (80.2\%) also contained a same-month wild-bird detection. Wild-bird counts followed recurring autumn--winter increases and summer troughs, whereas mammalian activity was more episodic and reached its largest national monthly peak in May 2024 (Fig.~\ref{fig:monthly}).

\begin{figure*}[htbp]
\centering
\includegraphics[width=0.92\textwidth]{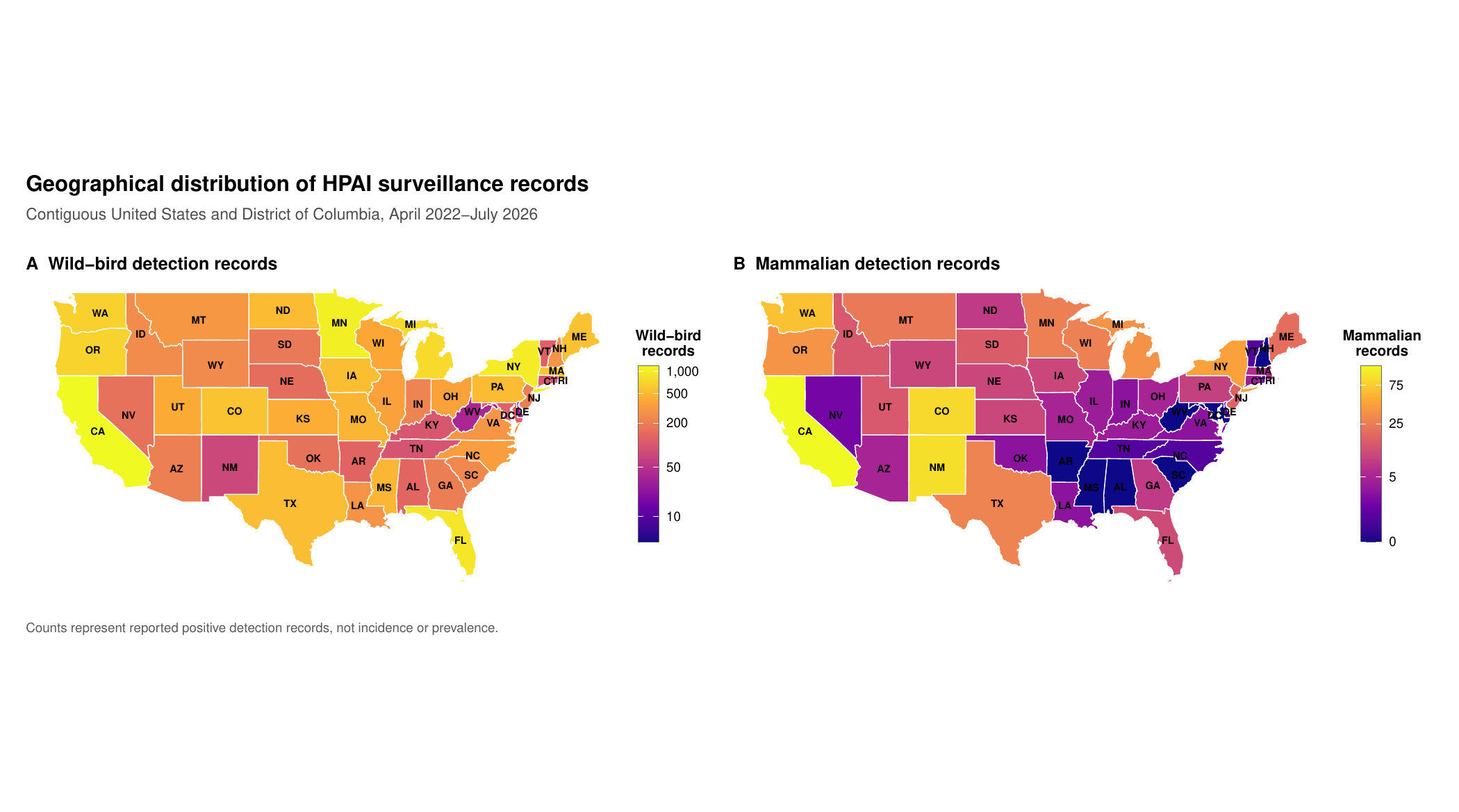}
\caption{Geographical distribution of wild-bird (A) and mammalian (B) HPAI surveillance records, April 2022--July 2026. Fill scales are log-transformed. Counts represent reported positive records; Alaska and Hawaii are not displayed.}\label{fig:maps}
\end{figure*}

The two streams also differed geographically. Wild-bird detections were broadly distributed, with the largest totals in California, Minnesota, New York and Florida. Mammalian records were more concentrated, with the largest totals in California, New Mexico, Colorado and Washington. Mammal-positive months showed that cumulative totals and temporal persistence were not identical: some states accumulated records in a small number of active months, whereas others reported detections over a longer period. The maps describe reporting distributions rather than state-level risks (Fig.~\ref{fig:maps}).

\subsection{Baseline distribution and zero reproduction}
All baseline candidate models used 2,601 observations and converged with positive-definite Hessians. NB1 had the lowest baseline AIC (2322.62; Akaike weight 0.998), followed by NB2 (2335.67), zero-inflated NB2 (2337.67) and Poisson (3696.10). The observed zero proportion was 0.8989. Poisson simulations produced a mean zero proportion of 0.8267 (95\% simulation range 0.7693--0.8764), which did not reproduce the observed sparsity. NB1, NB2 and zero-inflated NB2 produced mean zero proportions of 0.9060, 0.9027 and 0.9037, respectively, with simulation intervals containing the observed value. Thus, the high zero frequency did not by itself require a structural-zero component.

Residual evidence qualified the baseline AIC ranking. NB1 showed residual overdispersion ($p=0.004$); NB2 removed the dispersion problem but showed a modest uniformity departure ($p=0.0337$); zero-inflated NB2 showed no formally significant departure but added a parameter without improving fit relative to NB2. NB1 was consequently used to select the lag window, after which the selected structure was compared under NB1 and NB2.

\subsection{Lag-window selection}
The lag-complete dataset contained 2,499 observations, and all eight candidate NB1 models converged. The lags 0--1 model had the lowest AIC (2113.24; Akaike weight 0.633). The 0--2 and 0--3 windows had AICs of 2115.05 and 2116.76 ($\Delta$AIC 1.81 and 3.52), while lag 0 alone had AIC 2125.28. Adding lag 1 to lag 0 therefore improved AIC by 12.05 units. Lag 1 alone was less competitive (AIC 2154.02), and single-lag models for lags 2 and 3 were unsupported.

Likelihood-ratio tests against the no-bird baseline supported lag 0 ($\chi^2=69.55$, $p<0.001$), lag 1 ($\chi^2=40.82$, $p<0.001$) and the joint 0--1 window ($\chi^2=83.60$, 2 df, $p<0.001$), but not lag 2 ($p=0.149$) or lag 3 ($p=0.204$). Adjacent log-transformed bird lags were moderately correlated ($r=0.568$ for lags 0 and 1), while correlations declined with separation ($r=0.308$ for lags 0 and 2; $r=0.116$ for lags 0 and 3). Adding lags 2 and 3 left the lag 0 and lag 1 coefficients almost unchanged and produced additional coefficients close to zero. The 0--1 window was retained on fit, stability and parsimony grounds.

\subsection{Primary NB2 associations}
For the selected 0--1 window, NB2 had substantially lower AIC than NB1 (2085.88 versus 2113.24; $\Delta$AIC 27.36), with equal observations and parameter counts. Selected NB1 retained overdispersion ($p=0.010$), whereas NB2 showed no residual non-uniformity ($p=0.377$), dispersion failure ($p=0.962$), zero-count problem ($p=0.554$), abnormal outlier frequency ($p=0.257$), or temporal autocorrelation ($p=0.974$).

\begin{figure*}[htbp]
\centering
\includegraphics[width=0.7\columnwidth]{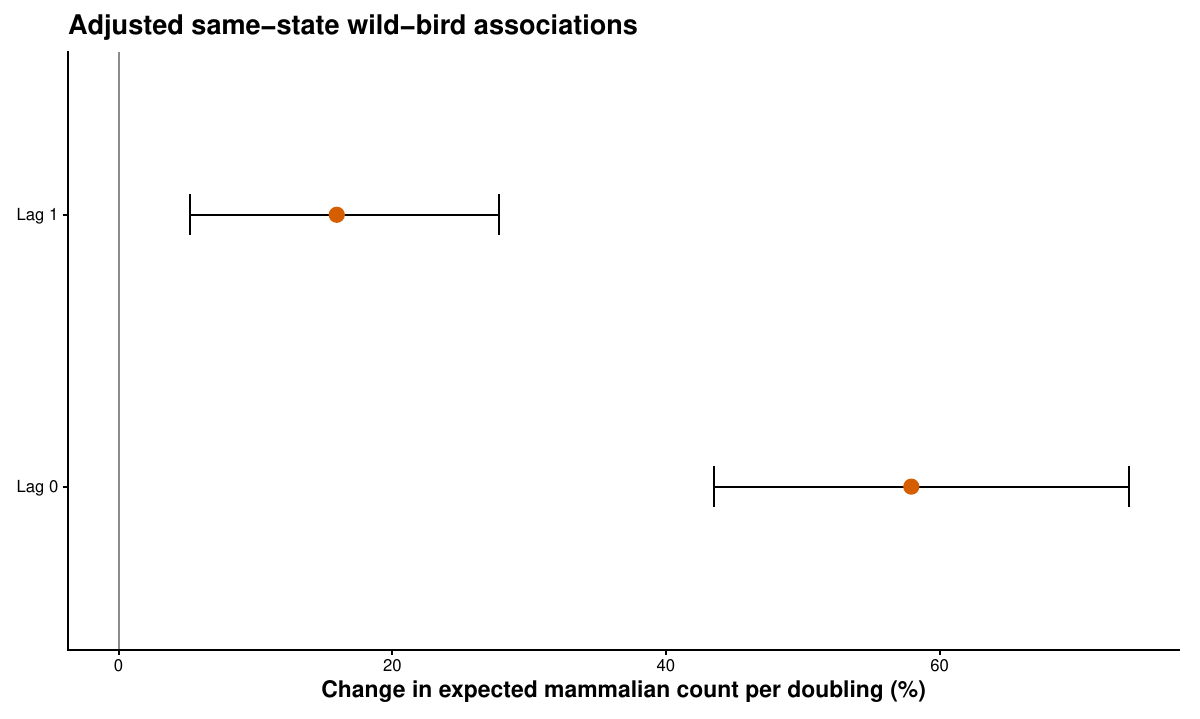}
\caption{Adjusted percentage changes in expected mammalian count per doubling of one plus same-state wild-bird exposure. Bars are 95\% confidence intervals.}\label{fig:effects}
\end{figure*}

In the NB2 model, doubling current-month bird exposure was associated with a 57.9\% increase in expected mammalian count, and doubling preceding-month exposure with a 15.9\% increase (Table~\ref{tab:effects}; Fig.~\ref{fig:effects}). The preceding mammalian count showed substantial persistence: doubling $1+M_{i,t-1}$ was associated with a 92.9\% increase. The cosine seasonal term was supported ($p<0.001$), while the sine term was weaker ($p=0.068$); the terms jointly represented annual seasonality.

\begin{table}[htbp]
\centering\footnotesize
\caption{Selected NB2 fixed effects.}\label{tab:effects}
\begin{tabularx}{\columnwidth}{@{}Xrr@{}}\toprule
Predictor & Change/doubling & 95\% CI\\\midrule
Birds, lag 0 & 57.9\% & 43.5--73.8\\
Birds, lag 1 & 15.9\% & 5.2--27.8\\
Mammals, lag 1 & 92.9\% & 53.6--142.3\\\bottomrule
\end{tabularx}
\end{table}

\subsection{Bordering-state exposure}
The contiguous-state analysis contained 2,401 observations, and all four candidate models converged. The same-state-only model had the lowest AIC (2013.92; Akaike weight 0.434). Adding neighbouring lag 1 yielded AIC 2014.75; adding neighbouring lag 0 yielded 2015.92; and adding both yielded 2016.49. Formal comparisons found no improvement for neighbouring lag 0 ($\chi^2=0.003$, $p=0.957$), lag 1 ($\chi^2=1.169$, $p=0.280$), or both jointly ($\chi^2=1.426$, $p=0.490$).

A doubling of neighbouring lag 0 exposure corresponded to a 0.4\% decrease (95\% CI: 13.8\% decrease to 15.1\% increase); neighbouring lag 1 corresponded to an 8.2\% increase (95\% CI: 6.2\% decrease to 24.8\% increase). Same-state lag 0 estimates remained between 57.6\% and 58.8\% across spatial extensions, and lag 1 estimates remained positive. The same-state model was retained.

\subsection{Temporally held-out prediction}
The training period contained 2,107 state-months and 634 mammalian records; the test period contained 294 state-months and 101 records. Adding wild-bird exposure reduced MAE from 0.774 to 0.590 (23.8\%), RMSE from 4.284 to 3.090 (27.9\%), and mean NLPD from 0.345 to 0.330 (4.3\%). The same-state model therefore improved both point prediction and the full predictive distribution, with a larger gain for point-error measures.

\begin{figure*}[htbp]
\centering
\includegraphics[width=0.6\textwidth]{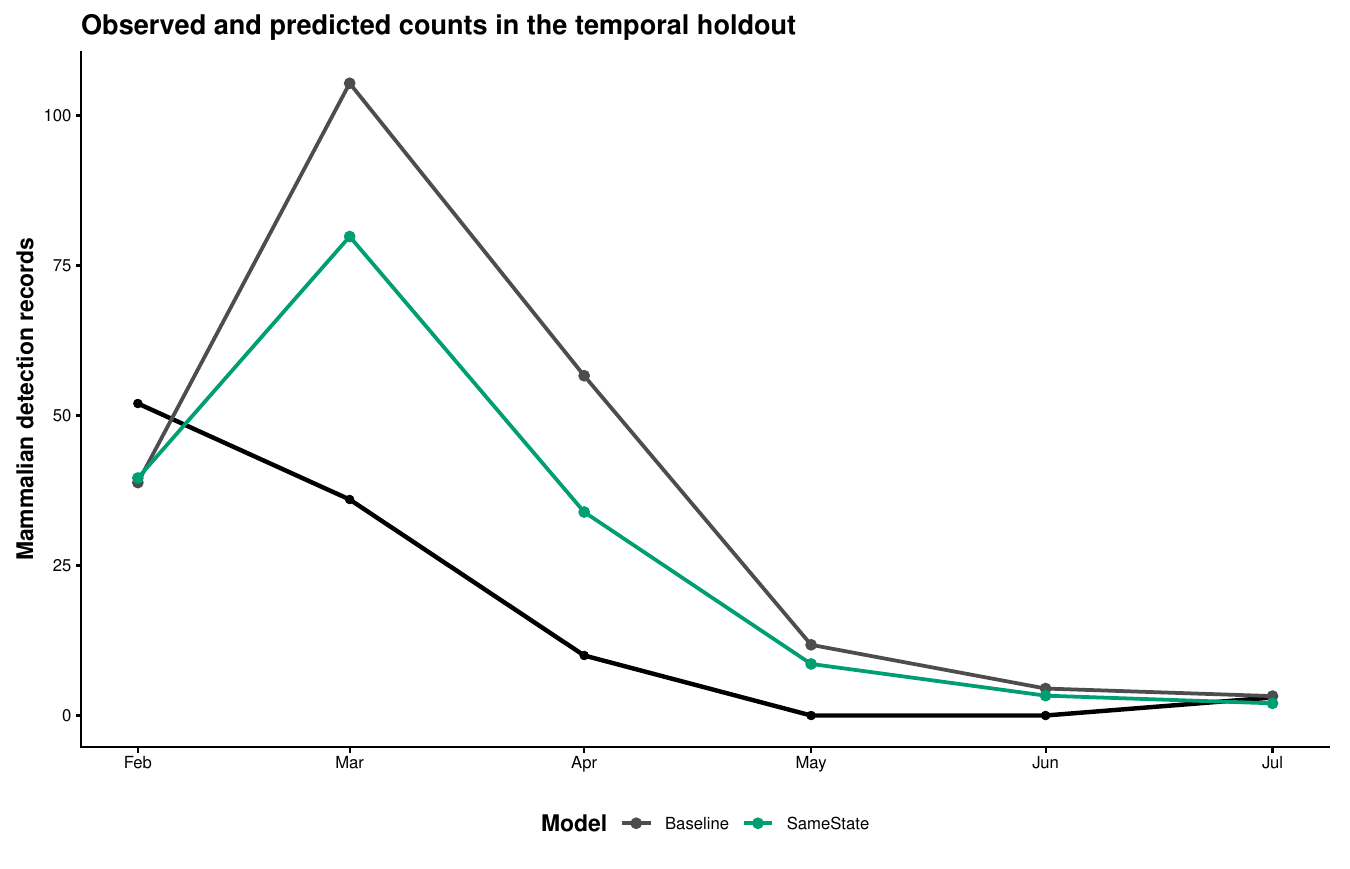}
\caption{Observed and predicted aggregate mammalian records in the temporal holdout. Predictions were conditional on the observed preceding mammalian count.}\label{fig:prediction}
\end{figure*}

Aggregate calibration remained imperfect. The baseline predicted 220.3 records and the same-state model 167.3, compared with 101 observed. The same-state model reduced aggregate overprediction by 53.0 records but still exceeded the observed total by 65.6\%. Empirical 80\% and 95\% coverage was 98.6\% and 99.3\%, indicating conservative intervals. Monthly results showed underprediction in February, overprediction in March and April, non-zero predictions when no records were observed in May and June, and close agreement in July (Fig.~\ref{fig:prediction}).

\begin{table*}[htbp]
\centering\footnotesize
\caption{Temporally held-out predictive performance, February--July 2026.}\label{tab:prediction}
\begin{tabular}{lrrrrrr}\toprule
Model & Predicted total & MAE & RMSE & Mean NLPD & 80\% coverage & 95\% coverage\\\midrule
Same-state wild-bird model & 167.3 & 0.590 & 3.090 & 0.330 & 98.6\% & 99.3\%\\
Autoregressive seasonal baseline & 220.3 & 0.774 & 4.284 & 0.345 & 97.3\% & 99.0\%\\\bottomrule
\end{tabular}
\end{table*}

\subsection{Exact-row-deduplicated sensitivity}
Deduplication retained 11,932 wild-bird and 609 mammalian primary-period records. The NB2 model converged with a positive-definite Hessian. The lag 0 effect was 59.7\% (95\% CI 43.6--77.6\%; $p<0.001$), compared with 57.9\% in the primary analysis; the lag 1 effect was 13.3\% (2.0--25.9\%; $p=0.020$), compared with 15.9\%. The preceding-mammal effect was 104.8\% (65.8--153.1\%). There was no residual overdispersion ($p=0.426$), zero-count problem ($p=0.556$), outlier-frequency problem ($p=1.000$), or temporal autocorrelation ($p=0.582$), although uniformity showed a modest departure ($p=0.031$). Exact repeated rows therefore did not explain the short-term bird associations.

\section{Discussion}
This study provides evidence that wild-bird HPAI surveillance records contain short-term and geographically local information about mammalian detection activity in the United States. Three findings are central. First, same-state wild-bird activity in the current and preceding month remained positively associated with mammalian counts after adjustment for annual seasonality, preceding mammalian activity and persistent state-level heterogeneity. Second, wild-bird activity in bordering states did not add detectable information once same-state exposure was included. Third, same-state wild-bird exposure improved temporally held-out point and probabilistic prediction, although the resulting predictions remained insufficiently calibrated for operational forecasting. Taken together, these findings support the use of wild-bird surveillance as a complementary indicator for targeting mammalian investigation, while not establishing the direction or mechanism of transmission.

\subsection{Interpretation of the short-term association}
The dominance of the contemporaneous association, accompanied by a smaller one-month-lagged effect, is consistent with a short period of epidemiological alignment between the two surveillance streams. Several mechanisms could generate this pattern. Mammals may encounter infected birds through predation or scavenging, or may be exposed to virus in water, soil or other contaminated substrates during periods of intense local wild-bird activity \cite{wood2010,root2020synanthropic,elsmo2023,enetwild2024mammals}. The expanding mammalian host range of clade 2.3.4.4b and documented infections in free-ranging terrestrial and marine mammals make these pathways biologically plausible \cite{puryear2024wildlife,plaza2024,peacock2025}. However, a contemporaneous surveillance association is not an incubation-period estimate. Wildlife mortality investigations, targeted testing and heightened awareness may increase sampling of birds and mammals at approximately the same time. Shared environmental conditions not fully represented by the annual harmonic terms may also affect both streams.

The positive coefficient for the preceding mammalian count was larger than the one-month bird effect, indicating substantial persistence within the mammalian surveillance series. This persistence may arise from continuing local events, repeated sampling during an investigation, delayed discovery of affected animals, repeated avian introduction, or transmission within a mammalian population. Epidemiological evidence consistent with mammal-to-mammal transmission has been reported during large H5N1 outbreaks in pinnipeds, while systematic reviews indicate that the extent and sustainability of such transmission remain context dependent \cite{enetwild2024mammals,peacock2025}. The present records cannot distinguish these mechanisms and should not be interpreted as reconstructing individual transmission chains.

The selected 0--1-month window was supported not only by AIC but also by likelihood-ratio tests, coefficient stability and the absence of evidence for additional effects at lags 2 and 3. This convergence across selection criteria reduces the likelihood that the retained window arose solely from one model-selection statistic. Nevertheless, monthly aggregation limits temporal resolution. A current-month association can include bird detections occurring before or after a mammalian record within the same calendar month. Analyses using exact collection dates, biologically specified exposure intervals and flexible distributed-lag functions could better resolve the order and duration of the association \cite{gasparrini2010distributed}.

\subsection{Local versus neighbouring-state information}
The failure of bordering-state exposure to improve model fit suggests that, at the state-month scale used here, the informative component of the wild-bird signal was predominantly local. This result is compatible with strong regional and host-specific structure in North American H5N1 activity \cite{damodaran2026,varga2026,canada2026spatiotemporal}. It may also reflect the way mammalian investigations are initiated: the probability of detecting an affected mammal is likely to respond more directly to wildlife events, laboratory submissions and veterinary awareness within the same administrative area than to surveillance activity averaged across several neighbouring states.

The null bordering-state result should not be interpreted as evidence that cross-border bird movement is epidemiologically unimportant. Shared political boundaries are an imperfect representation of migratory flyways, hydrological connections, habitat continuity and directional animal movement. Row-standardisation also gives each neighbouring state equal weight and may dilute a strong signal from one ecologically connected neighbour by averaging it with less relevant states. Future spatial models could replace binary adjacency with distance-decay weights, migration networks, watershed connectivity, species-specific flyways or viral genomic similarity. County-level modelling may also identify local cross-border relationships that are obscured by state aggregation.

\subsection{Predictive and surveillance implications}
Wild-bird exposure reduced MAE by 23.8\%, RMSE by 27.9\% and mean NLPD by 4.3\% relative to the autoregressive seasonal baseline. The larger improvement in RMSE indicates that bird information reduced some larger prediction errors, while the smaller NLPD improvement shows that gains in the full predictive distribution were more modest. These findings extend previous work demonstrating the potential of wild-bird and multi-source surveillance for HPAI early warning \cite{hill2023,shemmingspayne2024,movahedinia2026}. Importantly, the comparison was temporally held out rather than based only on in-sample fit, providing a more stringent assessment of incremental surveillance value \cite{tashman2000forecast}.

Predictive improvement did not imply adequate calibration. The same-state model predicted 167.3 records when 101 were observed, and empirical coverage exceeded the nominal 80\% and 95\% levels. The model therefore remained biased upward in aggregate and produced intervals that were too broad. Part of this discrepancy may reflect the unusual decline in mammalian detections during the holdout period, changes in surveillance intensity, or a relationship between bird and mammalian records that varied over time. Because predictions were conditioned on observed preceding mammalian counts, the evaluation represents repeated one-month-ahead prediction and not an autonomous six-month forecast. Operational implementation would require prospective updating, repeated validation across multiple seasons, calibration monitoring, and clear decision thresholds linked to the cost and capacity of additional mammalian sampling.

The practical implication is consequently one of prioritisation rather than prediction of individual spillover events. Elevated same-state wild-bird activity could be used as one component of a surveillance trigger, alongside mammalian morbidity or mortality reports, host ecology, poultry or livestock outbreaks, genomic information and measures of sampling effort. Such a trigger might justify intensified carcass collection, targeted testing of susceptible mammalian species, communication with wildlife rehabilitators and veterinary laboratories, or enhanced environmental investigation. This interpretation is consistent with contemporary calls for integrated animal-health surveillance and One Health preparedness \cite{koopmans2024,woah2026}.

\subsection{Robustness and limitations}
The principal estimates were similar after exact-row deduplication. The current-month effect changed from 57.9\% to 59.7\%, and the preceding-month effect from 15.9\% to 13.3\%. This stability indicates that exact repeated rows did not drive the primary conclusion. The selected NB2 model also reproduced the abundance of zeros without requiring a zero-inflated component and showed acceptable residual dispersion, zero-count behaviour, outlier frequency and temporal autocorrelation. These diagnostics strengthen confidence in the chosen mean-variance structure, although they do not address bias in the underlying surveillance process.

Several limitations remain important. First, the datasets contained positive detection records but not consistent numbers tested, negative results, host-population denominators or measures of surveillance effort. Counts therefore combine infection occurrence with the probability that an animal was encountered, submitted, tested and reported. Second, passive mortality surveillance, targeted studies and opportunistic submissions probably differed among states, host groups and periods. State random intercepts account for stable differences but not time-varying surveillance intensity. Third, exact-row deduplication represents two extreme assumptions: retaining all rows may preserve repeated specimens or animals but also duplicates, whereas removing exact rows may collapse genuinely distinct detections with identical recorded fields. Unique specimen and event identifiers would resolve this ambiguity.

Fourth, state-month aggregation may conceal county-level exposure pathways, species-specific relationships and short temporal ordering. Fifth, the neighbouring-state analysis did not incorporate migration direction, distance, habitat or sequence similarity. Sixth, the temporal holdout covered only one six-month period and included relatively few mammalian records; performance may differ during another epidemic phase. Finally, observational associations can remain confounded by unmeasured seasonality, environmental conditions, host abundance, reporting behaviour and coordinated surveillance. The estimated coefficients should not be interpreted as causal effects of wild-bird infection on mammalian incidence or as probabilities of spillover.

\subsection{Future research}
Further work should integrate testing denominators and surveillance effort wherever available, distinguish passive from targeted sampling, and model detection and infection processes separately. Finer spatial and temporal resolution could be combined with species traits, land cover, water bodies, weather, migratory connectivity and viral sequences. Hierarchical distributed-lag models could allow the bird--mammal relationship to vary by region, mammalian host group or epidemic phase. Repeated rolling-origin validation would provide a more reliable estimate of forecasting performance than a single holdout. Record-based early-warning models could also be linked to mechanistic models that represent within-herd transmission, asymptomatic infection and environmental reservoirs, while retaining a clear distinction between peer-reviewed evidence and emerging preprint results \cite{tiwari2025stochastic}. These extensions would help determine whether the present record-level association can be translated into a calibrated and operationally useful surveillance tool.

\section{Conclusion}
Wild-bird HPAI detection records provided independent short-term information about mammalian detection activity in the United States. The association was concentrated within the same state and within the current and immediately preceding month, persisted after exact-row deduplication, and improved prediction during a temporally held-out period. In contrast, wild-bird exposure averaged across bordering states did not improve model performance beyond the same-state signal.

The findings support using elevated local wild-bird activity as one indicator for prioritising mammalian surveillance. They do not demonstrate direct bird-to-mammal transmission, quantify spillover probability, or establish a forecasting system ready for operational use. Before implementation, the model requires prospective validation, explicit measures of surveillance effort, improved spatial representation and recalibration across multiple epidemic periods. Within those limits, integrating wild-bird and mammalian records offers a practical framework for strengthening coordinated wildlife and veterinary surveillance.

\section*{Acknowledgements} 
The author gratefully acknowledges the support provided through the AUT School of Engineering, Computer and Mathematical Sciences 2025 Early Career Research Award.

\section*{Ethics statement}
Public surveillance records were used; no experimental animal or human procedures were conducted.

\section*{Declaration of competing interest}
The author declares no competing interests.

\section*{Data availability}
Original data are publicly available from USDA APHIS \cite{usda2026}.


\begin{thebibliography}{99}

\bibitem{elsmo2023} E. J. Elsmo, A. W{\"u}nschmann, K. B. Beckmen, L. E. Broughton-Neiswanger, E. L. Buckles, J. Ellis, et al., ``Highly pathogenic avian influenza A(H5N1) virus clade 2.3.4.4b infections in wild terrestrial mammals, United States, 2022,'' \textit{Emerging Infectious Diseases}, vol. 29, no. 12, pp. 2451--2460, 2023. \url{https://doi.org/10.3201/eid2912.230464}. 

\bibitem{plaza2024} P. I. Plaza, V. Gamarra-Toledo, J. Rodr{\'\i}guez Eugu{\'\i}, and S. A. Lambertucci, ``Recent changes in patterns of mammal infection with highly pathogenic avian influenza A(H5N1) virus worldwide,'' \textit{Emerging Infectious Diseases}, vol. 30, pp. 444--452, 2024. \url{https://doi.org/10.3201/eid3003.231098}. 

\bibitem{peacock2025} T. P. Peacock, L. Moncla, G. Dudas, et al., ``The global H5N1 influenza panzootic in mammals,'' \textit{Nature}, vol. 637, pp. 304--313, 2025. \url{https://doi.org/10.1038/s41586-024-08054-z}. 

\bibitem{koopmans2024} M. P. G. Koopmans, B. C. Behravesh, A. A. Cunningham, et al., ``The panzootic spread of highly pathogenic avian influenza H5N1 sublineage 2.3.4.4b,'' \textit{The Lancet Infectious Diseases}, vol. 24, pp. e774--e781, 2024. \url{https://doi.org/10.1016/S1473-3099(24)00438-9}. 

\bibitem{damodaran2026} L. Damodaran, A. S. Jaeger, and L. H. Moncla, ``Ecology and spread of the North American H5N1 epizootic,'' \textit{Nature}, vol. 649, pp. 432--441, 2026. \url{https://doi.org/10.1038/s41586-025-09737-x}. 

\bibitem{wood2010} J. P. Wood, Y. W. Choi, D. J. Chappie, J. V. Rogers, and J. Z. Kaye, ``Environmental persistence of highly pathogenic avian influenza H5N1,'' \textit{Environmental Science \& Technology}, vol. 44, pp. 7515--7520, 2010. 

\bibitem{root2020synanthropic} J. J. Root and S. A. Shriner, ``Avian influenza A virus associations in wild, terrestrial mammals: A review of potential synanthropic vectors to poultry facilities,'' \textit{Viruses}, vol. 12, p. 1352, 2020. \url{https://doi.org/10.3390/v12121352}. 

\bibitem{puryear2024wildlife} W. B. Puryear and J. A. Runstadler, ``High-pathogenicity avian influenza in wildlife: A changing disease dynamic that is expanding in wild birds and having an increasing impact on a growing number of mammals,'' \textit{Journal of the American Veterinary Medical Association}, vol. 262, no. 5, 2024. \url{https://doi.org/10.2460/javma.24.01.0053}.

\bibitem{uhart2024elephantseals} M. M. Uhart, R. E. T. Vanstreels, M. I. Nelson, et al., ``Epidemiological data of an influenza A/H5N1 outbreak in elephant seals in Argentina indicates mammal-to-mammal transmission,'' \textit{Nature Communications}, vol. 15, p. 9516, 2024. \url{https://doi.org/10.1038/s41467-024-53766-5}. 

\bibitem{enetwild2024mammals} ENETWILD Consortium, O. Flavia, K. Sasha, et al., ``The role of mammals in avian influenza: A review,'' \textit{EFSA Supporting Publications}, vol. 21, no. 3, p. EN-8692, 2024. \url{https://doi.org/10.2903/sp.efsa.2024.EN-8692}. 

\bibitem{barbachano2023spillover} A. Barbachano-Guerrero, D. R. Perez, and S. L. Sawyer, ``How avian influenza viruses spill over to mammals,'' \textit{eLife}, vol. 12, p. e86051, 2023. \url{https://doi.org/10.7554/eLife.86051}. 

\bibitem{shemmingspayne2024} W. Shemmings-Payne, D. De Silva, C. J. Warren, et al., ``Repeatability and reproducibility of hunter-harvest sampling for avian influenza virus surveillance in Great Britain,'' \textit{Research in Veterinary Science}, vol. 173, p. 105279, 2024. \url{https://doi.org/10.1016/j.rvsc.2024.105279}.

\bibitem{stanislawek2024} W. L. Stanislawek, T. Tana, T. G. Rawdon, et al., ``Avian influenza viruses in New Zealand wild birds, with an emphasis on H5 and H7,'' \textit{PLOS ONE}, vol. 19, p. e0303756, 2024.

\bibitem{hill2023} L. Jung Kj{\ae}r, M. P. Ward, A. E. Boklund, et al., ``Using surveillance data for early warning modelling of highly pathogenic avian influenza in Europe reveals a seasonal shift in transmission, 2016--2022,'' \textit{Scientific Reports}, vol. 13, 2023. \url{https://doi.org/10.1038/s41598-023-42660-7}. 

\bibitem{varga2026} C. Varga, ``Spatiotemporal clustering of HPAI H5N1 at the wild waterfowl--poultry interface: Vector-specific spillover risks in the U.S., 2022--2025,'' \textit{PLOS ONE}, vol. 21, p. e0345354, 2026. \url{https://doi.org/10.1371/journal.pone.0345354}. \bibitem{movahedinia2026} Z. Movahedi Nia, N. Bragazzi, I. Gizo, et al., ``Integrating deep-learning methods and web-based data sources for surveillance, forecasting and early warning of avian influenza,'' \textit{Journal of the Royal Society Interface}, vol. 23, p. 20250578, 2026. \url{https://doi.org/10.1098/rsif.2025.0578}. 

\bibitem{wang2026} M. Wang, E. K. E. Laison, T. Philippsen, et al., ``Mechanistic modelling of highly pathogenic avian influenza: A scoping review revealing critical gaps in cross-species transmission models,'' \textit{PLOS ONE}, vol. 21, p. e0347929, 2026.

\bibitem{canada2026spatiotemporal} H. O. Fatoyinbo and H. Jeong, ``Spatiotemporal dynamics and ecological risk factors of high pathogenic avian influenza A(H5N1) in Canadian wildlife: A One Health surveillance analysis,'' \textit{Comparative Immunology, Microbiology and Infectious Diseases}, vol. 129, p. 102499, 2026. \url{https://doi.org/10.1016/j.cimid.2026.102499}.

\bibitem{tiwari2025stochastic} P. Tiwari, M. Smitha, and H. O. Fatoyinbo, ``Stochastic modelling and analysis of within-farm highly pathogenic avian influenza dynamics in dairy cattle,'' \textit{arXiv}, arXiv:2512.15118, 2025. \url{https://doi.org/10.48550/arXiv.2512.15118}. 

\bibitem{bolker2009glmm} B. M. Bolker, M. E. Brooks, C. J. Clark, S. W. Geange, J. R. Poulsen, M. H. H. Stevens, and J. S. S. White, ``Generalized linear mixed models: A practical guide for ecology and evolution,'' \textit{Trends in Ecology \& Evolution}, vol. 24, no. 3, pp. 127--135, 2009. \url{https://doi.org/10.1016/j.tree.2008.10.008}.

\bibitem{burnham2002model} K. P. Burnham and D. R. Anderson, \textit{Model Selection and Multimodel Inference: A Practical Information-Theoretic Approach}, 2nd ed. New York: Springer, 2002. \url{https://doi.org/10.1007/b97636}. 

\bibitem{warton2005zeros} D. I. Warton, ``Many zeros does not mean zero inflation: Comparing the goodness-of-fit of parametric models to multivariate abundance data,'' \textit{Environmetrics}, vol. 16, pp. 275--289, 2005. \url{https://doi.org/10.1002/env.702}. 

\bibitem{brooks2017} M. E. Brooks, K. Kristensen, K. J. van Benthem, A. Magnusson, C. W. Berg, A. Nielsen, H. J. Skaug, M. M{\"a}chler, and B. M. Bolker, ``glmmTMB balances speed and flexibility among packages for zero-inflated generalized linear mixed modeling,'' \textit{The R Journal}, vol. 9, no. 2, pp. 378--400, 2017. \url{https://doi.org/10.32614/RJ-2017-066}. 

\bibitem{hartig2026dharma} F. Hartig, \textit{DHARMa: Residual Diagnostics for Hierarchical (Multi-Level/Mixed) Regression Models}, R package version 0.5.0, 2026. Available: \url{https://CRAN.R-project.org/package=DHARMa}. 

\bibitem{gasparrini2010distributed} A. Gasparrini, B. Armstrong, and M. G. Kenward, ``Distributed lag non-linear models,'' \textit{Statistics in Medicine}, vol. 29, pp. 2224--2234, 2010. \url{https://doi.org/10.1002/sim.3940}. 

\bibitem{anselin1988spatial} L. Anselin, \textit{Spatial Econometrics: Methods and Models}. Dordrecht: Kluwer Academic Publishers, 1988. \url{https://doi.org/10.1007/978-94-015-7799-1}.

\bibitem{tashman2000forecast} L. J. Tashman, ``Out-of-sample tests of forecasting accuracy: An analysis and review,'' \textit{International Journal of Forecasting}, vol. 16, pp. 437--450, 2000. \url{https://doi.org/10.1016/S0169-2070(00)00065-0}. 

\bibitem{gneiting2007scoring} T. Gneiting and A. E. Raftery, ``Strictly proper scoring rules, prediction, and estimation,'' \textit{Journal of the American Statistical Association}, vol. 102, pp. 359--378, 2007. \url{https://doi.org/10.1198/016214506000001437}. 

\bibitem{gneiting2007calibration} T. Gneiting, F. Balabdaoui, and A. E. Raftery, ``Probabilistic forecasts, calibration and sharpness,'' \textit{Journal of the Royal Statistical Society: Series B}, vol. 69, pp. 243--268, 2007. \url{https://doi.org/10.1111/j.1467-9868.2007.00587.x}. 

\bibitem{usda2026} United States Department of Agriculture, Animal and Plant Health Inspection Service, ``Detections of highly pathogenic avian influenza in mammals,'' USDA APHIS, 2026. Available: \url{https://www.aphis.usda.gov/livestock-poultry-disease/avian/avian-influenza/hpai-detections/mammals?page=1}. Accessed: Oct. 4, 2026.

\bibitem{woah2026} FAO, WHO, and WOAH, ``Updated joint public health assessment of recent high pathogenicity avian influenza A(H5) virus events in animals and people,'' May 18, 2026. Available: \url{https://www.who.int/publications/m/item/updated-joint-fao-who-woah-public-health-assessment-of-recent-high-pathogenicity-avian-influenza-a(h5)-virus-events-in-animals-and-people}. 

\end{thebibliography}

\end{document}